\documentclass[%
prx,longbibliography,twocolumn,
superscriptaddress,
 amsmath,amssymb,
 aps,
]{revtex4-1}
\usepackage[utf8]{inputenc}
\usepackage{graphicx}
\usepackage{xcolor}
\usepackage{siunitx}
\usepackage{hyperref}
\usepackage[normalem]{ulem}
\newcommand{\FigSq}{Supplementary Figure~S1}
\newcommand{\FigNCptpcP}{Supplementary Figure~S2}
\newcommand{\FigHistRg}{Supplementary Figure~S3}
\newcommand{\FigLossTangent}{Supplementary Figure~S4}
\newcommand{\FigClusters}{Supplementary Figure~S5}
\newcommand{\FigIsoPerco}{Supplementary Figure~S6}

\usepackage{pdfpages}
\makeatletter
\AtBeginDocument{\let\LS@rot\@undefined}
\makeatother
\usepackage{pgffor}

\begin{document}
\title{Gelation as condensation frustrated by hydrodynamics and mechanical isostaticity}

\author{Hideyo Tsurusawa}
\thanks{These authors contributed equally to this work}
\affiliation{ {Department of Fundamental Engineering, Institute of Industrial Science, University of Tokyo, 4-6-1 Komaba, Meguro-ku, Tokyo 153-8505, Japan} }
\author{Mathieu Leocmach}
\thanks{These authors contributed equally to this work}
\affiliation{Univ Lyon, Université Claude Bernard Lyon 1, CNRS, Institut Lumière Matière, F-69622, Villeurbanne, France}
\author{John Russo}
\affiliation{ {School of Mathematics, University of Bristol, Bristol BS8 1TW, United Kingdom} }
\author{Hajime Tanaka}
\email{tanaka@iis.u-tokyo.ac.jp}
\affiliation{ {Institute of Industrial Science, University of Tokyo, 4-6-1 Komaba, Meguro-ku, Tokyo 153-8505, Japan} }

\begin{abstract}
Colloidal gels have unique mechanical and transport properties that stem from their bicontinous nature, in which a colloidal network is intertwined with a viscous solvent, and have found numerous applications in foods, cosmetics, construction materials, and for medical applications, such as cartilage replacements. So far, our understanding of the process of colloidal gelation is limited to long-time dynamical effects, where gelation is viewed as a phase separation process interrupted by the glass transition. However, this picture neglects two important effects: the influence of hydrodynamic interactions, and the emergence of mechanical stability. With confocal microscopy experiments, here we successfully follow the entire process of gelation with a single-particle resolution, yielding time-resolved measures of internal stress and viscoelasticity from the very beginning of the aggregation process. First, we demonstrate that the incompressible nature of a liquid component constrains the initial stage of phase separation, assisting the formation of a percolated network. Then we show that this network is not mechanically stable and undergoes rearrangements driven by self-generated internal stress. Finally, we show that mechanical metastability is reached only after percolation of locally isostatic environments, proving isostaticity a necessary condition for the stability and load bearing ability of gels rather than the glass transition. Our work reveals the crucial roles of momentum conservation in gelation in addition to the conventional purely out-of-equilibrium thermodynamic picture.
\end{abstract}

\maketitle

\section{Introduction}

A gel is a soft solid composed of two intertwined phases: a solid network and a liquid solvent. 
They are an ubiquitous state of matter in every-day life, making up most of the foods we eat, the cosmetics we use, concretes, and our own organs. In colloidal gels, the network is composed of colloidal particles bonded together by attractive forces. 
Such colloidal assemblies are out of equilibrium, as the thermodynamic ground state of the system involves the macroscopic separation between a particle-rich and a particle-poor phase. 
Despite the thermodynamic driving force towards compactness, the gel persists due to the dynamical arrest of the network, often described as a glass transition~\cite{Piazza1994,verduin1995phase,Verhaegh1996,Tanaka1999VPScolloid,trappe2001jamming,poon2002,Cardinaux2007,lu2008gelation}.
This has led to the popular physical picture that a gel is formed by dynamical arrest of bicontinuous spinodal decomposition due to glass transition. 
The direct link between spontaneous gelation and spinodal decomposition has been carefully confirmed by combining experiments and theories~\cite{lu2008gelation}. 
This recently established scenario is certainly a large step towards a more complete  understanding of colloidal gelation. 

However, this picture still leaves some fundamental problems unanswered: (i) The knowledge of ordinary spinodal decomposition predicts that the minority colloid-rich phase should form isolated clusters rather than the observed percolated network~\cite{onuki2002phase}.
(ii)  A colloidal gel is sometimes formed by a network made of thin arms, which are too thin to be regarded as glasses. 
This casts some doubt on the popular scenario of dynamic arrest due to a glass transition. Indeed, the glass transition is defined as a kinetic transition and has no direct link to mechanical stability in a strict sense. Slow dynamics and mechanical stability are conceptually different. In an extreme case, for example, a gel formed by bonds with a short  lifetime can be ergodic and in an equilibrium state. 
(iii) A gel often displays superdiffusive behavior, detected as the compressed exponential decay of a density correlation function, during ageing, as observed by time-resolved spectroscopy techniques~\cite{Cipelletti2000,Ramos2001,Ruta2014a} and recently simulations~\cite{Tanaka2007,Bouzid2016,Chaudhuri2017}. The origin of this phenomenon and its relation to problem (ii) are still elusive. 

Several mechanisms have been proposed to try to rationalize some of these issues. Fluid momentum conservation can play an important role in phase separation, giving to hydrodynamics an active role in structural evolution~\cite{Tanaka1999VPScolloid,Tanaka2000VPS}.
There have been some numerical studies on the role of hydrodynamics~\cite{Furukawa2010,Varga2015a} and mechanics~\cite{Cipelletti2000,Tanaka2007,Colombo2014,Testard2014,Zia2014,Chaudhuri2017} in colloidal gelation, however, there has so far been no experimental access to these problems due to the lack of a method to follow the whole kinetic processes with single-particle resolution in both space and time. 
More importantly, some questions on the emergence of elasticity, which is the most fundamental physical property of gels, have still remained unanswered. 
The stability of gels is ascribed to the formation of locally favored structures, or local energy-miminum configurations~\cite{Royall2008c}, while the mechanics of the network is being recognized to play a major role in the ageing behavior of gels~\cite{Bouchaud2002,Cipelletti2003,Tanaka2007,Bouzid2016}.
Purely geometrical conditions for mechanical stability have also been proposed. \citet{Kohl2016} have found that in dilute suspensions, a final gel state could be obtained only for interaction potentials where directed percolation was observed. \citet{Hsiao2012} have found that strain-induced yielding coincide with the loss of rigid clusters. Rigidity was defined using Maxwell criterion for isostaticity, that is 6 neighbors per particle~\cite{Maxwell1864}.
However, the relationship between local structures, dynamic arrest, and the emergence of elasticity remains poorly understood even at a fundamental level. 

In order to address these problems experimentally, in this Article we study roles of hydrodynamics and mechanics in colloidal gelation  by dynamical confocal microscopy observation of the entire gelation process, from the beginning to the end with particle-level resolution. To reveal the roles of momentum conservation in gelation, it is crucial to develop a special protocol to initiate phase separation without causing any harmful flow (see Methods). 
We stress that such flow not only makes the initial state ill-defined, but also has permanent influence on the kinetic pathway, as well as the final state.   
This protocol allows us to experimentally elucidate  
(a) the importance of hydrodynamic interactions in forming and stabilizing elongated clusters and their percolation, and 
(b) the crucial role of percolation of isostatic structures in conferring mechanical stability to the network. 
These findings shed new light on the mechanisms of gel formation and coarsening, and also on the fate of gels.

\section{Single-particle phase separation}

\subsection{System design}
\begin{figure}
	\includegraphics[width=8.0cm]{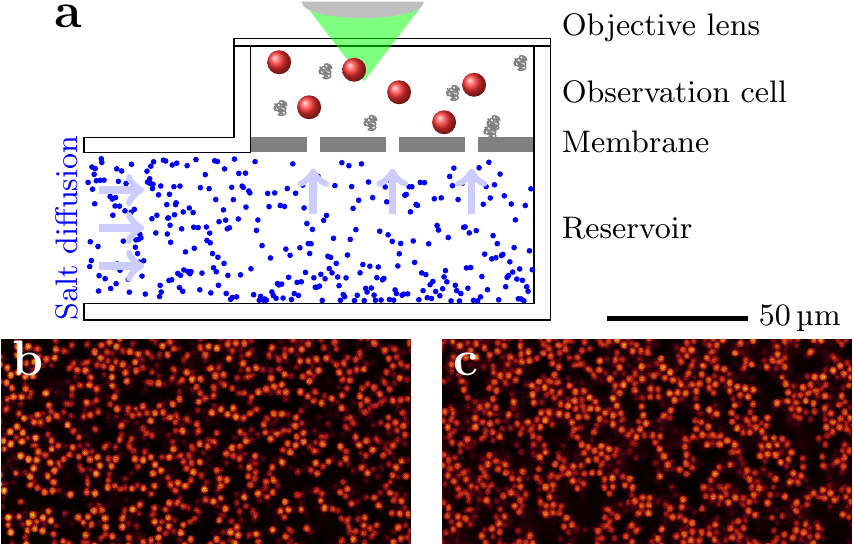}
	\caption{Reservoir cell. (a) Sketch of our experimental setup. The observation cell contains initially colloids, polymer and no salt. (b) Confocal slice of a gel formed \textit{in situ} by our method ($\phi=25.5~\%$, $c_p=1.4$ mg/g), 1 hour after gelation. (c) Idem for a gel at the same state point formed \textit{ex situ} and immediately pumped into a capillary. 
	}
	\label{fig:cell_vs_cap}
\end{figure}

\begin{figure*}
	\includegraphics{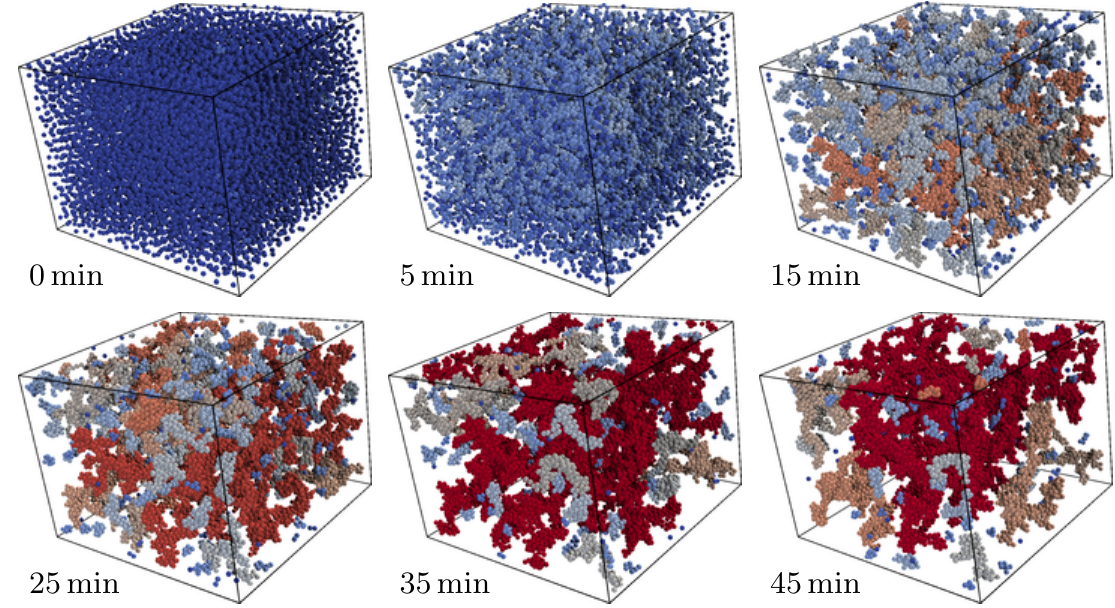}
	\caption{Snapshots from the entire gelation process reconstructed via particle-tracking of a typical sample close to the cluster-gel line ($\phi=7.5~\%$, $c_p=1$ mg/g) using the salt-injection protocol. Particles are colored according to the size of the cluster they belong to, going from blue for monomers to red for the percolated cluster. 
	}
	\label{fig:image}
\end{figure*}
In our protocol, we first enclose a salt-free suspension of sterically and charge-stabilized colloids and non adsorbing polymers in a thin microscopy cell sketched in Fig.~\ref{fig:cell_vs_cap}(a). 
The bottom wall of the cell is an osmotic membrane providing contact with a long channel full of the same solvent mixture. 
Salt dissolution and subsequent migration of the ions along the channel and through the membrane induce screening of the electrostatic repulsion, revealing the depletion potential well due to the polymers. 
The time needed for the ions to diffuse from the membrane across the cell thickness is of the same order of magnitude as the Brownian time of the particles $\tau_B=\SI{10}{\second}$. 
This relation between the two key timescales enables us to switch instantaneously (physically) from a long-range repulsive to a short-range attractive system without any external solvent flow, which has never been achieved experimentally before.
This causes uniform gelation starting from the homogeneous Wigner crystal state, allowing \textit{in situ} confocal microscopy observation throughout the process from a well-defined initial time.

Figure~\ref{fig:cell_vs_cap}(b) and (c) compare the final structures of two gels prepared at the same state point with the two different protocols: the first one by our salt-injection protocol, and the latter by the conventional approach, where a gel is formed in a capillary and then shear melted at the start of the experiment. 
Already a visual inspection reveals that the latter is coarser, highlighting that shaking or shear melting protocols~\cite{lu2008gelation,Bartlett2012} are not equivalent to a quench. 
Our special quench protocol provides an ideal experimental platform to make a comparison with theory and simulations. 
However, we note that Brownian Dynamics simulations cannot reproduce our experimental results even with the same quench, because they neglect the solvent mediated hydrodynamic interactions~\cite{Furukawa2010,Varga2015a}.

In Fig.~\ref{fig:image}, we show a computer reconstruction from experimental coordinates of a typical gelation experiment at a relatively low volume fraction ($\phi=7.5\%$). 
Immediately before ions enter the cell ($t=0$), the suspension is in a Wigner crystal state where the particles are far apart due to long range Coulomb repulsion. 
As soon as the charges are screened, the particles begin to aggregate and form clusters that progressively connect to each other while coarsening to finally percolate over the field of view around $t=\SI{35}{\minute}$, see Methods. 
The absence of macroscopic flow can be confirmed in Supplementary Movie 1.

\subsection{Phase separation dynamics}
\begin{figure}
\centering
\includegraphics{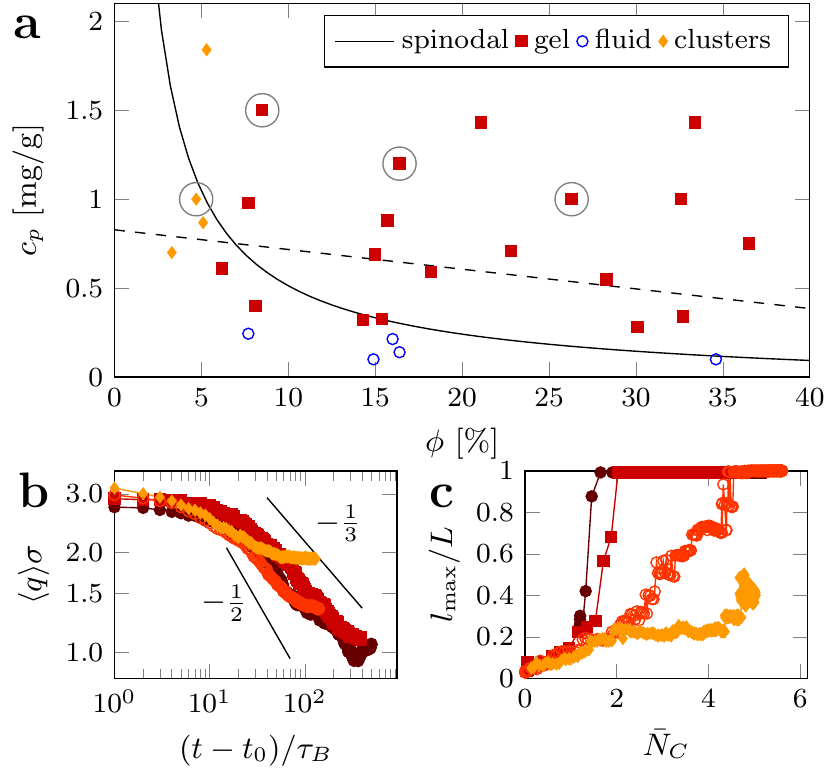}
\caption{Different regimes of gelation.  
(a) Phase diagram. 
Experimental points are categorised based on the final state obtained in the reservoir cell. 
The spinodal line is obtained from free volume theory in polymer dilute regime. The dashed line is the polymer overlap concentration in the free volume. 
State points analysed in b and c are circled.
(b) Growth of the characteristic wave number. 
By increasing density: $\phi=4.2,\,8,\,16,\,27~\%$, $c_p=1,\,1.5,\,1.2,\, 1$ mg/g for 
\textcolor{red!40!yellow}
{$\blacklozenge$}, 
\textcolor{red!80!yellow}
{$\circ$}, 
\textcolor{red!80!black}
{\tiny$\blacksquare$} and 
\textcolor{red!40!black}
{$\bullet$} respectively.
The lines are possible scaling laws for the intermediate coarsening regime. 
(c) Comparison of system evolution in terms of largest cluster extent and of mean coordination number for the same samples.}
\label{fig:phasediag}
\end{figure}

In Fig.~\ref{fig:phasediag}(a), we show the phase diagram, where we can divide the state points into three regions based on the final state obtained by our protocol: 
at low polymer concentration ($c_p<0.2$ mg/g) a sample fully relaxes to a fluid state; 
at very low colloid volume fraction ($\phi<0.05$) and high polymer concentration the particles condense into long-lived well separated clusters as observed in~\cite{Lu2006}; 
in the rest of the explored phase space we observe a long-lived space spanning network. 
In the phase diagram we also plot the spinodal (continuous) and the polymer overlap concentration (dashed) lines as obtained from free-volume theory calculations~\cite{Fleer2008}. Despite the limitations of the theory, the agreement between the spinodal line and our experiments is rather satisfactory, with the only exception being the region of small colloidal volume fractions and high polymer concentration (see, e.g., Ref. ~\cite{lu2008gelation}).

To confirm whether the different samples follow spinodal-decomposition kinetics, we compute the time dependent static structure factor $S(q,t)$. 
For all gel and cluster samples, we observe the appearance of a low $q$ peak in $S(q)$, see \FigSq, which is characteristic of spinodal decomposition in a system with a conserved order parameter.
Then, we compute the characteristic wave number $\langle q \rangle$ and show its temporal evolution in Fig.~\ref{fig:phasediag}(b) for various colloidal volume fractions. 
The curves for all samples follow a master curve coherent with spinodal decomposition kinetics: At short times $\langle q \rangle(t)$ shows a plateau indicating that the low $q$ peak builds up at a constant wave number corresponding to distances of about $2\sigma$. 
This plateau is characteristic of the early stage spinodal decomposition, which is described by Cahn's linear theory~\cite{onuki2002phase}. 
At intermediate times, on the other hand, we observe coarsening with $\langle q \rangle \sim t^{-\alpha}$, with an exponent $\alpha$ which is compatible with both $\alpha=1/3$, typical of spinodal decomposition without dynamical asymmetry between the two phases, and $\alpha=1/2$, which is often observed in viscoelastic phase separation (see, e.g., Ref.~\cite{Furukawa2010}).
Due to the narrow range of this power law regime and finite size effects, we cannot conclude definitely on the exponent value.
Finally at longer times each sample deviates from the master curve to form a plateau indicating arrest.
The more dilute samples arrest sooner, but reciprocal space information does not allow to identify whether the origin of arrest is different between clusters and percolated networks.

To identify the origin of the dynamical arrest observed above, we now take advantage of the single-particle level resolution of our reconstructed trajectories. 
To characterise this path, we compute the instantaneous mean number of neighbors $\bar{N}_C$, or coordination number, that quantifies the compactness of the structure. 
We also compute the spatial extent of the largest cluster $l_\text{max}$ that we normalize by the size of the field of view $L$ to obtain a measure of the distance to percolation of the system.
Figure~\ref{fig:phasediag}(c) shows a system trajectory in the $(l_\text{max}/L, \bar{N}_C)$ plane for various colloidal volume fractions. 
All trajectories show a linear increase of both cluster size $l_\text{max}/L$ and number of neighbors $\bar{N}_C$ at early times, corresponding to the first plateau in Fig.~\ref{fig:phasediag}(b). This is followed by the coarsening stage, which happens differently depending on the density. 
At high densities, coarsening occurs after percolation, which happens within the first few $\tau_B$ after charge screening by salt. 
At low densities, percolation never takes place and coarsening results in the compaction of individual clusters, that keeps their overall size $l_\text{max}/L$, while increasing the number of neighbors $\bar{N}_C$. 
We did not observe any Ostwald ripening among clusters, indicating that the diffusive evaporation-condensation coarsening mechanism is negligible compared to cluster collisions and coalescence, as expected for colloidal viscoelastic phase separation~\cite{Tanaka2000VPS}.
At intermediate densities, we observe the process detailed in Fig.~\ref{fig:image}: formation of low-compacity clusters that then slowly connect together to build the percolating network.
This process can take hundreds of $\tau_B$ and is competing with cluster compaction, as indicated by the oblique trajectory (red open circles) in Fig.~\ref{fig:phasediag}(c).
Particle-level quantities thus demonstrate that the path to gelation is not universal and depends on the colloid volume fraction even within the gelation region.
By contrast, polymer concentration, i.e. the depth of the attraction potential, has little effect on the path to gelation, see \FigNCptpcP{}.

Our observations indicate that both the cluster and gel phases are due to viscoelastic spinodal decomposition~\cite{Tanaka1999VPScolloid,Tanaka2000VPS}: network-type spinodal for the gel, and droplet-type spinodal for the clusters, where strong dynamical asymmetry between colloids and the solvent leads to unique roles of hydrodynamics and mechanics in phase separation.

In section~\ref{sec:hydro} we will explore the role of hydrodynamics. To do so, we will restrict to dilute cases where percolation occurs late or never, leaving enough time to observe disconnected clusters. In section~\ref{sec:mech}, we will explore the precise mechanism of arrest and the emergence of mechanical rigidity by studying the dynamics within the network of percolating samples.

\section{Role of hydrodynamics}
\label{sec:hydro}
\begin{figure}
	\centering
	\includegraphics{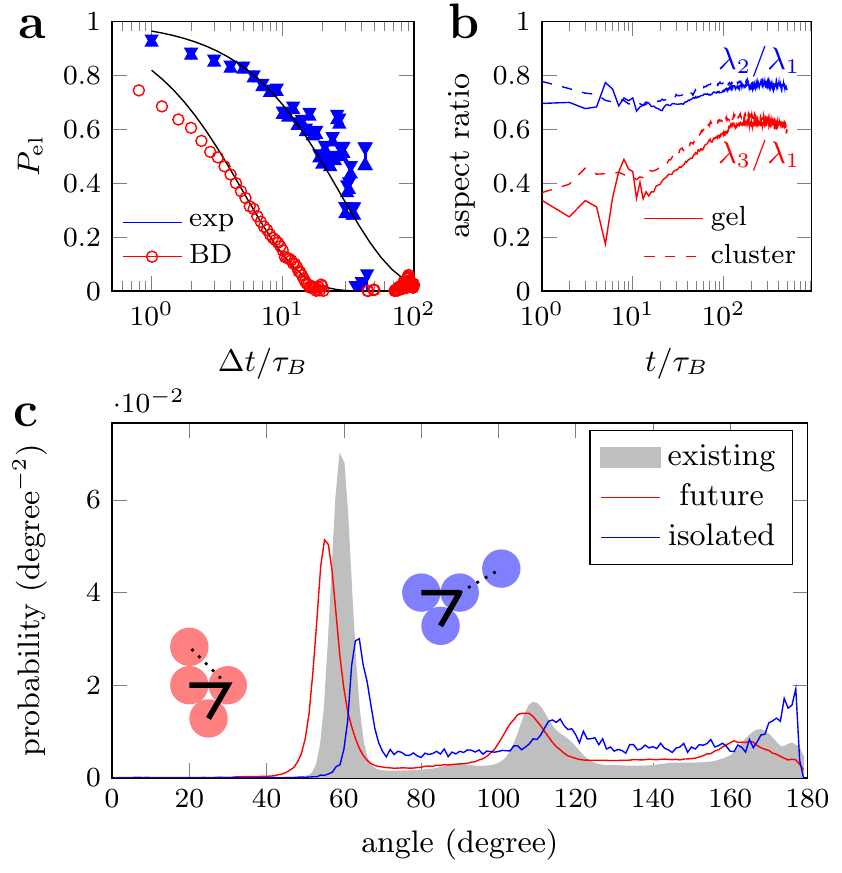}
	\caption{Hydrodynamics. (a) Probability of staying elongated for a triplet in a non-percolating sample ($\phi=4~\%$, $c_p=1$ mg/g, blue) and in corresponding BD simulations. The continuous lines are the respective best exponential fits of characteristic time $27\tau_B$ and $5\tau_B$ respectively. (b) Evolution of the aspect ratios of clusters of 4 particles and more in the same sample (dashed lines) and in a percolating sample ($\phi=8~\%$, $c_p=1.5$ mg/g, continuous lines). (c) Bond angle distribution relative to existing bonds (grey), to a future bond (red) or to a future bond involving an isolated particle (blue) obtained in the percolating sample. Future bonds are shifted to smaller angles, whereas gas adsorption takes place from larger angles. Insets sketch both cases, with present bonds drawn thick and future bonds drawn dotted.}
	\label{fig:hydro}
\end{figure}

In Fig.~\ref{fig:hydro}(a), we follow the compaction of clusters made of only three particles in a non-percolating sample. 
The time-averaged probability distribution of the radius of gyration $R_g$ of these triplets shows two peaks on both sides of $R_g^*=0.8\sigma$, see \FigHistRg. 
For $R_g<R_g^*$ the cluster is compact, with a structure close to an equilateral triangle. 
For $R_g>R_g^*$ the three particles are aligned and the cluster is elongated. 
We found that just after the quench triplets have a slightly higher probability of being elongated. Afterwards, they either connect to other clusters or relax to the more stable compact state. 
To follow this relaxation, we define the probability to stay elongated as 
\begin{equation}\label{eq:elongation}
P_\text{el}(\Delta t) = \left\langle P\left(\delta_i(t+\Delta t)|\delta_i(t)\right)\right\rangle_{t,i}, 
\end{equation}
where $\delta_i(t)$ is the probability for the triplet $i$ to be elongated at time $t$. 
Figure~\ref{fig:hydro}(a) (blue line) shows that the decay of $P_\text{el}(\Delta t)$ is exponential with a characteristic time of $27\tau_B$. In the same figure we also plot
(red symbols) the same quantity computed from Brownian Dynamics simulations of short-range attractive colloids designed to match the experiments (see Methods), in which the triplet compaction process is simulated in absence of hydrodynamic interactions. For the  simulations we observe a considerably faster exponential decay compared to the experiments, suggesting that the triplet compaction process is indeed slowed down significantly by hydrodynamic interactions.

The shape of clusters composed of more than 3 particles cannot be followed in the same way. 
Instead, we compute the principal moments of gyration of individual clusters $\lambda_j$, ordered such that $\lambda_1\geq\lambda_2\geq\lambda_3$, with
use the aspect ratios $\lambda_2/\lambda_1$ and $\lambda_3/\lambda_1$ to quantify the departure from sphericity.  
In Fig.~\ref{fig:hydro}(b), we show the evolution of the average value of these aspect ratios either for a non-percolating sample (dashed line), or before percolation for a percolating one (continuous line). 
In both cases, we observe that the clusters are originally not compact and become more isotropic over tens of $\tau_B$. 
As can be seen in Fig.~\ref{fig:image} and \FigClusters, structural isotropy is recovered only after the fusion of many anisotropic clusters into a branched structure that may or may not be percolating.

These observations can be understood as due to hydrodynamic effects. 
Indeed in a solvent, particles cannot converge freely to form compact structures~\cite{Tanaka1999VPScolloid,Furukawa2010}. 
The compaction is delayed by the incompressibility of the solvent, which allows only divergence-free transverse flow fields.  
Furthermore, clusters influenced by hydrodynamic interactions tend to be more elongated, less compact. 
We can test this hypothesis by measuring at which angle particles meet relative to existing neighbors.
If influenced by hydrodynamics, particles should avoid the direction of existing neighbors and come from more open angles.
In Fig.~\ref{fig:hydro}(c) we show the bond angle distribution for three different sets of bonds: 
(i) existing bonds,
(ii) bonds that will form within the next $\tau_B$ (\emph{future bonds}), 
(iii) future bonds where the newly attached particle is a monomer. 
As expected, existing bonds (i) are preferentially at a 60$^\circ$ angle, indicating stable packing, with secondary peaks coherent with a mixture of tetrahedral and hexagonal packing. 
Future bonds (ii) have more acute angles and almost never 180$^\circ$, since they are mostly due to particles attached to second neighbors, see sketch in Fig.~\ref{fig:hydro}(c). 
Here hydrodynamics plays no role. 
By contrast, future bonds (iii) involving isolated particles form at more obtuse angles, with a clear peak around 180$^\circ$. 
This confirms that hydrodynamics has a significant influence on particle aggregation and explains why clusters are initially elongated.

Consequently, long-lived elongated clusters have a higher probability to meet via either rotational or translational diffusion than compact spherical clusters. 
Hydrodynamics explains why in a rather dilute regime we can observe immediate formation of elongated clusters and then their slow, hydrodynamically-assisted aggregation into a percolated structure.
We stress that this is a direct consequence of large-size disparity between colloidal particles and solvent molecules, which leads to the physical situation where discrete solid objects are floating in a continuum liquid.

\section{Emergence of mechanical stability}
\label{sec:mech}

\begin{figure}[t]
\includegraphics{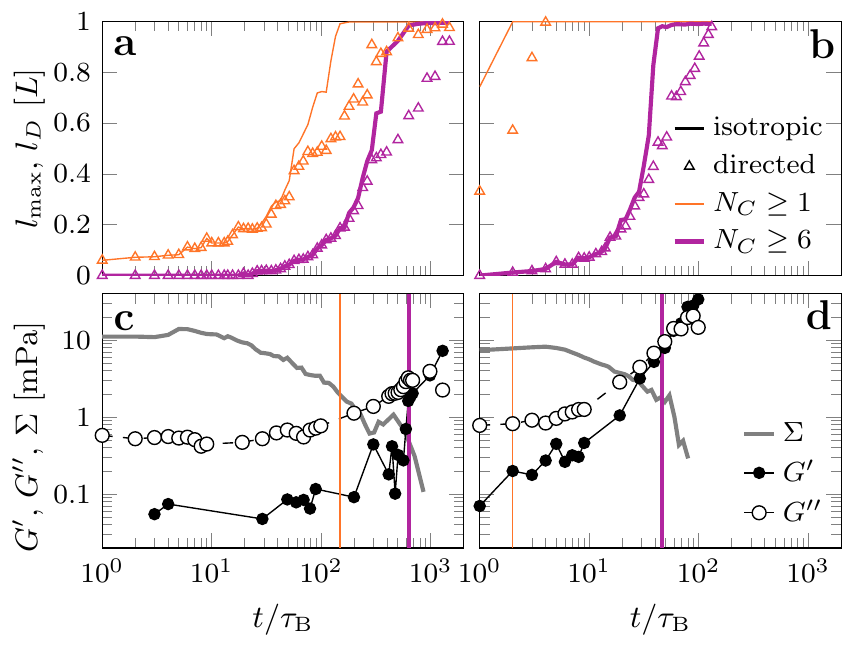}
\caption{Evolution of space-spanning microstructure and mechanical response. 
(a) and (b): Percolation processes, for a dilute $(\phi=8~\%,\,c_p=\SI{1.5}{mg/g})$ 
and a dense $(\phi=27~\%,\,c_p=\SI{1}{mg/g})$ sample.
The processes of isotropic and directed percolation of all particles ($N_C \geq 1$) are respectively plotted as thin orange curve and orange symbols.
The processes of isotropic and directed percolation of isostatic particles ($N_C \geq 6$) are respectively plotted as 
thick purple curve and purple symbols.
(c) and (d): Mechanical response for the same samples. 
Elastic ($G^\prime$) and viscous ($G^{\prime\prime}$) shear moduli at high frequency ($f=0.1\tau_\mathrm{B}^{-1}$), obtained by two-particle microrheology, are drawn respectively as filled and open circles.
The thick grey curve is the internal stress $\Sigma$ obtained from the measure of bond-breaking probability. 
The thin orange (left) and thick purple vertical lines show the isotropic percolation times for all particles ($t_\text{perco}$) and isostatic particles ($t_6$) respectively.
}
\label{fig:mecha_vs_perco}
\end{figure}

\subsection{Percolation}
To better grasp the timing of space-spanning microstructural changes in dilute and dense regime, we plot in Fig.~\ref{fig:mecha_vs_perco}(a) and (b) the time evolution of largest connected cluster $l_\text{max}$ (orange curves), and define the \emph{isotropic percolation} time ($t_\text{perco}$) as the moment when $l_\text{max}>0.95L$. In dense case, $l_\text{max}$ reaches the size $L$ of the observation window in a few Brownian times, whereas it takes hundreds of $\tau_B$ in the dilute case. 
We also display the maximum spatial extent of directed paths $l_D$ (orange symbols in Fig.~\ref{fig:mecha_vs_perco}(a) and (b)).
A directed path is defined as a path with no loop or turning back, such that every step is in either the positive X, Y, or Z directions. We thus define the \emph{directed percolation} time ($t_D$) as the moment when $l_D>0.95L$. 
Finally, in Fig.~\ref{fig:mecha_vs_perco}(a) and (b) we also consider \emph{isostatic clusters}, defined as clusters which comprise only particles that have at least six bonded neighbors. For isostatic clusters we then plot both $l_\text{max}$ (purple curves) and $l_D$ (purple symbols) (see also Supplementary Movie 2). We define $t_6$ as the time of percolation of isostatic clusters. 
We observe that directed percolation of all particles (at $t_D$, see orange triangles) and isotropic percolation of isostatic particles (at $t_6$, see thick purple curve) occur simultaneously in the dilute regime. However the two time scales are well separated in the dense regime. 
This separation of time scales offers the opportunity to test the role of both type of space spanning microstructures in the mechanical stability of gels.

\subsection{Mechanical stability and percolations}

The solid nature of a material is most often defined from linear mechanical response. 
However for colloidal gels, mechanical stability cannot be predicted without an understanding of internal stresses~\cite{Bouzid2016}.
Here we are able to extract both information from our particle-level experiments.
We use the particles themselves as passive microrheological probes to extract the elastic ($G^\prime$) and viscous ($G^{\prime\prime}$) parts of the shear modulus, see Methods.
We also extract the average value of the internal stress $\Sigma$ from the bond breaking rate, see Methods.
Results are shown in Fig.~\ref{fig:mecha_vs_perco}(c) and (d) for direct comparisons with the microstructure.

The typical ranges of stresses and moduli we measure extend below \SI{0.1}{\milli\pascal}, well below sensitivity of conventional rheometers.
That is why previous microscopic studies on the mechanics of colloidal gels have been restricted to the comparison of the structure before and after a large amplitude shear flow with no simultaneous measure of the stress response~\cite{Smith2007,Lindstrom2012,Hsiao2012}.
From Fig.~\ref{fig:mecha_vs_perco}(c) and (d) we see that, as expected, all samples are purely viscous at short times, with a value of $G^{\prime\prime}$ consistent with the viscosity of a hard sphere suspension at their respective volume fraction.
Internal stresses are high at short time, reflecting the stretching due to hydrodynamic frustration.
The emergence of mechanical stability is captured simultaneously from both the linear viscoelasticity measurements, with the crossing between $G^\prime$ and $G^{\prime\prime}$, and the internal stress, which exhibits a sharp drop in $\Sigma$.
The timing of the emergence of elasticity is thus unambiguous and occurs well after isotropic percolation time $t_\text{perco}$ (see orange vertical lines in Fig.~\ref{fig:mecha_vs_perco}(c) and (d)).
This generalizes observations by \citet{Kohl2016} on the final state of dilute samples.

In dilute samples, the elastic behavior occurs in the same time scale as directed percolation of all particles.
However, isotropic percolation of isostaticity also occurs simultaneously. 
Therefore we have to look at the dense regime to disentangle the two possible microstructural causes. 
Indeed in the dense regime the elastic behavior emerges well after directed percolation of all particles, in the same time scale as isotropic percolation of isostaticity.
This allows to lay the main result of this article: the emergence of rigidity is caused by isotropic percolation of isostatic clusters, able to bear stress across the sample. 

\subsection{Directed or isostaticity percolations}

\begin{figure}
\includegraphics{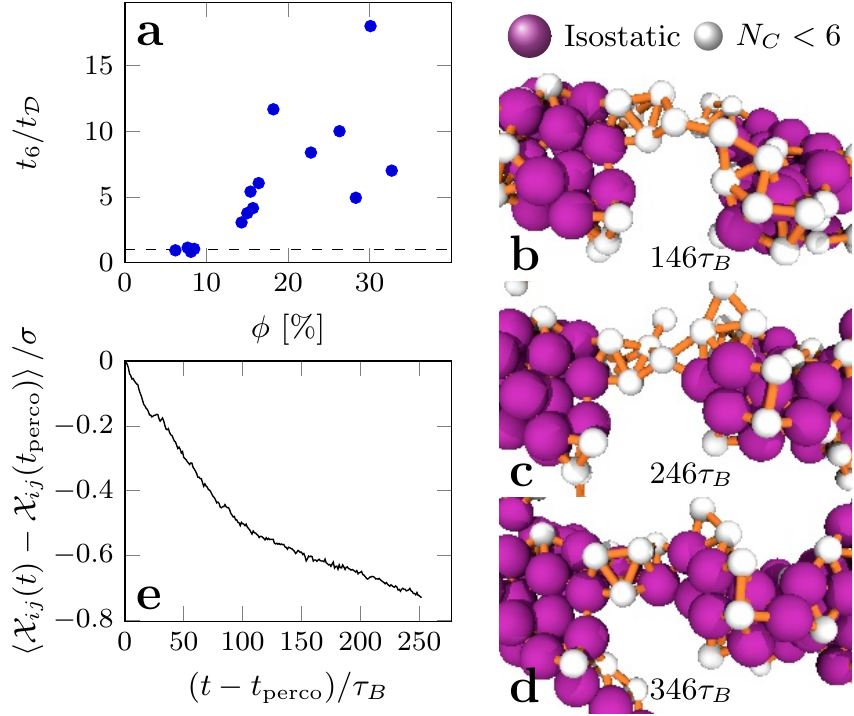}
\caption{Directed percolation and isostaticity percolation. 
(a) Ratio of the time of isostaticity percolation by the time to directed percolation, function of colloid volume fraction. Horizontal dashed line shows when both times are equal. 
(b) Detail of a reconstruction from confocal coordinates at percolation time in a dilute sample $(\phi=8~\%,\,c_p=\SI{1.5}{mg/g})$. 
Isostatic particles are drawn to scale, non isostatic ones are drawn smaller for clarity. 
The bond network is displayed in orange. 
(c) and (d) Same as (b) at later times.
(e) Increment of Euclidian distance between two isostatic clusters, averaged over all such pairs initially connected by a non-isostatic network strand. The reference time is the percolation time.}
\label{fig:directed_isostatic}
\end{figure}

In Fig.~\ref{fig:directed_isostatic}(a), we compare across all our experiments the time to percolation of isostaticity $t_6$ to the time to directed percolation $t_\mathrm{D}$. 
We confirm that at high volume fractions, typically $\phi>14\,\%$, the two phenomena are decoupled, with $2<t_6/t_\mathrm{D}<20$ depending on the state point.
By contrast at lower $\phi$, both percolations occur simultaneously, independently of the attraction strength. 

The reason for this coincidence can be understood by the specific path to gelation in the dilute regime. We have seen in Fig.~\ref{fig:phasediag}(c) that isotropic percolation occurred at a late stage, when the clusters had time to compact. Indeed, Fig.~\ref{fig:mecha_vs_perco}(a) shows that at percolation time, isostatic particles already form clusters that reach up to a tenth of the observation window. Cluster-size distribution (see \FigIsoPerco) and three dimensional reconstruction in Fig.~\ref{fig:directed_isostatic}(b) show that these isostatic clusters are compact, typically 3 to 5 particles in diameter and linked by non-isostatic bridges. The floppiness of these bridges prevents directed paths to reach percolation.

From this situation, percolation of isostaticity proceeds by the compaction of the floppy bridges. Importantly, this compaction takes place without adsorption of new particles onto the bridge. Compaction is a local process that involves no particle migration but only creation of new bonds, as shown in Fig.~\ref{fig:directed_isostatic}(b)-(d). 
Consistently, this compaction leads to a straightening and a shortening of the strands. We quantify this shortening by computing the Euclidean distance $\mathcal{X}_{ij}(t)$ between the centers of mass of two isostatic clusters $i$ and $j$. The increment of this distance with respect to percolation time, averaged over all cluster pairs connected by a floppy bridge is shown on Fig.~\ref{fig:directed_isostatic}(e). The observed shortening is about $0.7\sigma$ or 25\% of the initial length. Directed percolation becomes possible when a percolating path has become straight enough, which implies isostaticity. That is why directed percolation and isostaticity percolation occur simultaneously in the dilute regime.

\subsection{Stress-induced network breakup}

\begin{figure}
\includegraphics[width=8.5cm]{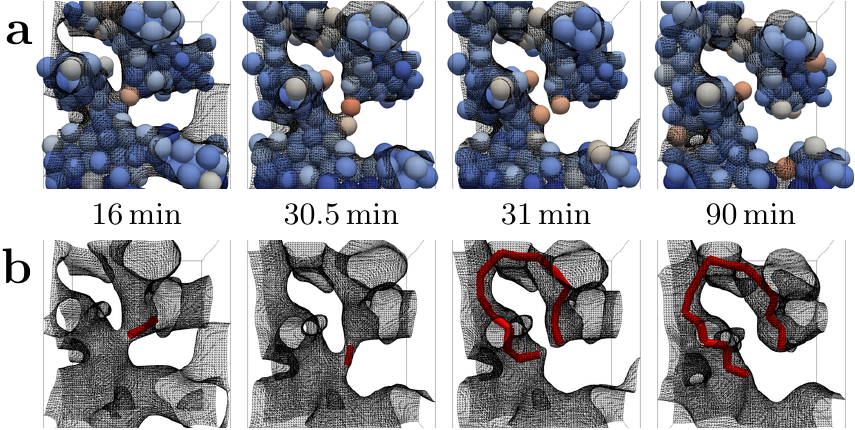}
\caption{Breakup of the network by internal stress. 
(a) Reconstruction from experimental coordinates ($\phi=29~\%$, $c_p=\SI{0.7}{mg/g}$) of strand rupture event. 
Particles are drawn to scale and colored by a measure of two-fold symmetry $q_2$ (see Sec.~\ref{sec:q2} on its definition) from blue (low) to red (high). We note that $q_2$ is a measure of the degree of local stretching. 
(b) Same event from a topological point of view. 
The red line indicates the shortest on-graph path between the two particles of interest, whose drastic change clearly indicates the breakup event. The meshed surface is a Gaussian coarse-graining of the network pattern.}
\label{fig:breakup}
\end{figure}

In a dense system, after directional percolation of all particles, the number of nearest neighbors monotonically increases to minimize the energy of the structure, resulting in the growth of isostatic configurations, as discussed above. 
During this process the mechanical tension internal to the network grows, driving it towards compaction, which can lead to network coarsening accompanying bond breakage (see Fig. \ref{fig:breakup}(a) and (b)). 
Unlike in simulations \cite{Tanaka2007}, we cannot directly measure the local internal stress at this moment, but we can still see its effects through the local stretching measured by the degree of two-fold symmetry $q_2$ (see the particle color in Fig.~\ref{fig:breakup}(a)). 
From this, we may say that a bond breakage event is the consequence of stress concentration on a weak bond, leading to local stretching of the bond, and its eventual breakup.  
In other words, mechanical stress acts against diffusive particle aggregation (or compaction), which is the stress-diffusion coupling characteristic of phase separation in dynamically asymmetric mixtures \cite{Tanaka1999VPScolloid,Tanaka2000VPS}. This stress-driven aging is accompanied by mechanical fracture 
of the percolated network structure by the self-generated mechanical stress. 
The mechanical stability can be attained only after the formation of a percolated isostatic structure, which is a necessary condition for a structure to be  mechanically stable. When the percolated isostatic structure can support the internal stress everywhere, the system can attain mechanical stability.  

\section{Discussion and conclusion}

In summary, we have observed with particle-level resolution the entire process of gelation at various state points for the first time. 
The early stages are characterized by the universal features of spinodal decomposition, with clusters emerging with a constant $q$ vector.
However, we have shown that hydrodynamic interactions hinder the formation of compact clusters, and give to the coarsening process a non-universal behavior. 
At high volume fractions, elongated structures immediately percolate into a thin, mechanically unstable network that undergoes stress-driven rearrangements enabling the formation of locally isostatic structures that finally percolate. 
At low volume fractions, percolation is delayed, thus initially elongated clusters have the time to compact before eventually connecting into a percolating structure.
Isostatic clusters thus already exist at percolation time but are linked by floppy strands that have to compact to induce isostaticity percolation.
In both cases, microrheology indicates that neither isotropic percolation nor directed percolation ensure mechanically solidity. 
Instead, we show that mechanical gelation is characterized by the isotropic percolation of isostatic structures. 

The picture of gelation that emerges from our observations is far more rich than previously understood, and suggest that mechanical stability plays a more important role than dynamical arrest.
Firstly, our study clearly shows that the spinodal decomposition in gelation is under a strong influence of mechanics.
Dynamic asymmetry between big colloidal particles and small solvent molecules leads to hydrodynamically assisted percolation and coarsening under self-generated mechanical stress. 
Secondly, unlike the glass transition, which is kinetically defined as the point above which the relaxation time is slower than the observation time, the arrest is due to percolation of isostaticity.  Thus, we argue that a key feature of gelation is {\it viscoelastic spinodal decomposition arrested by isostaticity percolation}. 
Then, the mechanical stability  of a gel is determined by a competition between the yield stress of the isostaticity network and the internal stress towards network shrinking produced by the interface free-energy cost. 
Since a gel is not in an equilibrium state and the stress can be concentrated in a weak part of \emph{the network}, perfect mechanical stability may never be attained, resulting in slow ageing via either surface diffusion or bond breakage.   

In contrast to a purely out-of-equilibrium thermodynamic picture of gelation, an understanding based on the mechanical equilibrium and isostaticity might pave the way to a more operative description of colloidal gels, and allow complex issues to be addressed in terms of mechanics and rheology.
For example, stress-driven ageing plays a fundamental role in the formation of porous crystals~\cite{tsurusawa2017formation}. The spontaneous delayed collapse of colloidal gels~\cite{Kamp2009,Bartlett2012} could be viewed as the final overcome of the mechanical frustration. 
Under small stresses also a delayed yielding is observed~\cite{Gibaud2010,Sprakel2011,Leocmach2014}. 
Despite a sustained attention, 
the yielding process of colloidal gels still lacks a general consensus. 
For instance we do not know why some colloidal gels display a yield stress fluid behavior, that is a reversible yielding and no fracture~\cite{Gibaud2010,Grenard2014}, whereas others display a brittle solid behavior with the irreversible opening of fractures~\cite{Leocmach2014}.
Understanding colloidal gels as a both non-ergodic and mechanically stabilised state of matter may help solving these issues. 

\section{Materials and Methods}

\subsection{Experimental}

\subsubsection{Samples}
We used colloidal particles made of \textsc{pmma} (poly(methyl methacrylate)) copolymerized with methacryloxypropyl terminated \textsc{pdms} (poly(dimethyl siloxane)) for steric stabilisation~\cite{klein2003}, with 2~\% of methacrylic acid to allow electrostatic repulsion, and with
(rhodamine isothiocyanate)-aminostyrene for fluorescent labelling~\cite{bosma2002}. Colloids are dispersed in a mixture of cis-decalin (Tokyo Kasei) and bromocyclohexane (Sigma-Aldrich) that matches both optical index and density of the colloids.

To induce short-ranged depletion attraction, we use \SI{8.4}{\mega\dalton} polystyrene (TOSOH) as non-adsorbing polymer. The radius of gyration in theta solvent is estimated to \SI{110}{\nano\metre}. Here the solvent may be regarded as ``good'' and a Flory scaling of the measurements described in Ref. ~\cite{lu2008gelation} yields $R_g=148$ nm. In the absence of salt, the Debye length is expected to reach several \si{\micro\metre} and the (weakly) charged colloids experience a long range electrostatic repulsion~\cite{Royall2003}. We confirm that colloids never come close enough to feel the short-ranged attraction and form a Wigner crystal.

\subsubsection{Special experimental protocol to initiate phase separation without harmful flow}
Gelation of micron-size colloids suitable for quantitative confocal microscopy is usually induced by depletion attractions due to polymers in the solvent.
The experimental protocols that have been used so far for studying the kinetics of phase separation and gelation are as follows: 
(1) Colloidal suspensions and polymer solutions are mixed just before an experiment, and after mixing transferred to a capillary tube as quickly as possible. 
(2) A mixture, which is already in a final state point in the phase diagram and intrinsically unstable, or phase-separated, is vigorously stirred just before an experiment to break pre-existing phase separated structures by shear melting. 
However, these protocols have two common serious deficiencies. 
Firstly the initial state can never been homogeneous perfectly, and so there already exist particle aggregates at $t=0$. 
Secondly, the mixing inevitably involves turbulent flow, which does not decay but remains when the observation is initiated. 
The gelation process observed by these conventional protocols inevitably suffers from the influence of ill-defined initial static and dynamic conditions,
and it has been almost impossible to access the very initial stage of gelation without interference of pre-existing aggregates and/or turbulent flow. 

We overcome these limitations as follows.
We use a colloidal system that is charge stabilised at long range, has a short range depletion attraction, and is also sterically stabilised causing nearly hard sphere repulsion at contact. 
We disperse colloidal particles and non-adsorbing polymers in a  mixture of organic solvents that matches both the refractive index and the density of the particles.
We realize a density matching of the order of $10^{-4}$ between the density of the colloids and of the solvent, enough to observe the late stage of gelation with little influence of gravity despite our large particle size (gravitational Peclet number $\mathrm{Pe}<10^{-6}$).
Because of the weakly polar nature of the solvent mixture (its dielectric constant $\epsilon_r = 5\sim6$), the Debye screening length is about $\kappa^{-1}=\SI{10}{\micro\metre}$, long enough for the large colloids (diameter $\sigma=\SI{2.75}{\micro\metre}$) to form a homogeneous Wigner crystal in the mixture~\cite{Klix2010a}. 
The short ranged ($\sim \sigma/10$), depletion attraction caused by the polymers is masked by the electrostatic repulsion.

The colloids and polymers are contained in an observation cell ($\SI{10}{\milli\metre\squared} \times \SI{200}{\micro\metre}$) made of glass in contact with an half-open glass channel approximately 400 times larger in volume, via a millipore filter with pore size of \SI{100}{\nano\metre} that allows the salt through but neither polymer nor colloid (see Fig.~\ref{fig:cell_vs_cap}(a)). The channel is filled with the same solvents at density matching composition. At the beginning of the experiment, solid tetrabutylammonium bromide (Fluka) is introduced to the channel. Data acquisition starts within \SI{30}{\second} after salt introduction. Our procedure induces practically no solvent flow in the observation cell. We confirmed the presence of undissolved salt several days after mixing, indicating that the observation cell was brought to saturation concentration.

Given the diffusion constants of Bromide and alkyl cation (6 and \SI{2e-10}{\metre\squared\per\second}~\cite{Campbell2005}), we estimate the characteristic diffusion time of salt from top to bottom of the order of \SI{10}{\second}. Therefore, we reach uniform final salt concentration into the observation cell within only a few Brownian times of the colloids. Indeed we measured a delay of about \SI{1}{\minute} between the aggregation at the bottom and at the top of the cell. We define the initiation time of the aggregation process when the maximum of the $g(r)$ jumps from the lattice constant of the Wigner crystal to the hard-core diameter $\sigma$.

We collect the data on a Leica SP5 confocal microscope, using \SI{532}{\nano\metre} laser excitation. The temperature was controlled on both stage and objective lens, allowing a more precise density matching. The scanned volume is at least $82 \times 82 \times \SI{85}{\micro\metre^3}$. The particle coordinates are tracked in three dimensions (3D) with an accuracy of around $0.03\sigma$~\cite{LeocmachColloids}.

\subsection{Simulations}
To simulate the process of triplet compaction in absence of hydrodynamic interactions, we use Langevin dynamic simulations, where the characteristic damping time of the velocities $\tau_D$ is chosen to be equal to the Brownian time $\tau_B$, i.e. the time it takes a colloid to diffuse its diameter. We use a generalised LJ potential (with exponent $n=100$ and interaction strength $\epsilon=8K_BT$) chosen to match the second virial coefficient of the Asakura-Osawa potential corresponding to experimental conditions (ratio of polymer to colloid diameter, $q=0.1$ and strength $\epsilon=8K_BT$). Following Ref.~\cite{lu2008gelation}, the process of matching the second virial coefficient should ensure equivalent dynamical behavior for all short-range potentials. The elongation probability Eq.~(\ref{eq:elongation}) is computed by running two hundred independent simulations and measuring the statistics of open and compact configurations of the triplets.

\subsection{Analysis}

\subsubsection{Characterisation of the system.} 
From direct confocal measurements~\cite{Royall2007, Poon2012}, we estimate the hard-core diameter of our colloids ($\sigma=\SI{2.75}{\micro\metre}$) and the range of the interaction potential (that confirmed our scaling of $R_g$ within $1\%$), leading to a polymer-colloid size ratio $q = 2R_g/\sigma = 0.10(6)$. Spinodal line  and polymer overlap concentration line on Fig.~\ref{fig:phasediag}(a) are calculated from this size ratio using the generalized free volume theory~\cite{Fleer2008}.

\subsubsection{Detection of bonds.}
In principle the attraction well of the depletion extends to $\sigma+2R_g$, however, resolution-dependent tracking imprecision and systematic errors do not give a precise enough estimate of such short distance. Therefore we consider two particles bonded when their distance is shorter the first minimum of $g(r)$, i.e. \SI{3.55}{\micro\metre}. This defines the bond graph that we analyse using NetworkX library~\cite{Hagberg2008}. We have checked that the precise choice of this distance does not affect significantly our results, in particular percolation times. We consider that a bond is effectively broken when it does not reform within $10\tau_B$.

\subsubsection{Estimation of the internal stress.}

\begin{figure}
\includegraphics{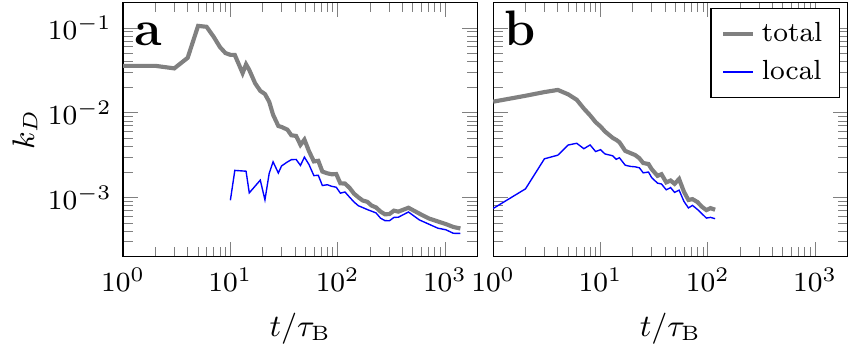}
\caption{Bond breaking rates. 
for a dilute (a) $(\phi=8~\%,\,c_p=\SI{1.5}{mg/g})$ and a dense (b) $(\phi=27~\%,\,c_p=\SI{1}{mg/g})$ percolating sample.
The thick grey curve shows the total breaking rate.
The thin blue curve counts only breaking events after which the two particles still have a common neighbor.
}
\label{fig:breaking_rate}
\end{figure}

To measure internal stresses, we model the bond breaking rate using a Kramers approach~\cite{Kramers1940}. In absence of force $\mathcal{F}$ acting on a bond, the dissociation rate is
\begin{equation}
k_D(\mathcal{F}=0) = \omega_0 \exp\left( -\frac{E_A}{k_\mathrm{B}T}\right),
\end{equation}
where $E_A$ is the depth of the potential and $\omega_0$ an attempt frequency that depends on the precise shape of the potential~\cite{Plischke2006} and on the diffusion constant in the depletion shell~\cite{Smith2007}. For small forces, the rate becomes~\cite{Lindstrom2012}
\begin{equation}
k_D(\mathcal{F}) = k_D(\mathcal{F}=0) \exp\left(\frac{\mathcal{F} \delta}{k_\mathrm{B}T}\right)
\end{equation}
with $\delta$ the width of the potential, here $2R_g$. If we can measure $k_D$ in absence of force, we can obtain the force at all times:
\begin{equation}
\mathcal{F}(t) = \frac{k_\mathrm{B}T}{\delta} \log\frac{k_D(t)}{k_D(\mathcal{F}=0)}.
\end{equation}
We convert the force into the internal stress $\Sigma$ using the area of contact between depletion shells $\Sigma = 2\mathcal{F} /(\pi\sigma\delta)$. Here we measure $k_D(\mathcal{F}=0)$ by supposing that at long times, once hydrodynamic stresses can be neglected, local rearrangements of the network are force-free, that is when the involved particles keep a common neighbor after bond breaking, the long time limit of the blue curve in Fig.~\ref{fig:breaking_rate}.

\subsubsection{Microrheological measurements of viscoelasticity.}
We perform two-particle microrheology following Ref.~\cite{Crocker2000}. Briefly, we compute the two-point mean square displacement $\langle\Delta r^2\rangle_D (t, \Delta t)$, averaged over all couples of particles $(i,j)$ so that $\sigma < r_{ij} < r_\text{max}$ and particle $i$ is further away than $r_\text{max}$ from any edge of the observation window.
We chose $r_\text{max}$ as the fourth of the shortest dimension of the observation window.
Using a generalised Stokes-Einstein relation, we obtain the complex modulus $G(\omega,t)$.

In general the crossing of $G^\prime$ and $G^{\prime\prime}$ depends on the frequency. 
That is why the gel point is defined as the time when $G^\prime$ and $G^{\prime\prime}$ both scale as identical power laws of frequency which corresponds to a loss tangent $G^{\prime\prime}/G^\prime$ independent of frequency~\cite{Chambon1987}.
In \FigLossTangent{} we shows the evolution of $G^{\prime\prime}/G^\prime$ in a percolating sample, for three frequencies. 
Gelation point occurs at the very end of the experimental time, slightly later than the crossing of $G^\prime$ and $G^{\prime\prime}$ at high frequency.
We confirmed this trend for all samples and we can safely conclude that in all cases the gel point occurs soon after percolation of isostaticity within the experimental window.


\subsubsection{Characterisation of the degree of local stretching.}
\label{sec:q2}
To detect local 2-fold symmetry and thus elongation, we use Steinhardt bond orientational order parameter~\cite{steinhardt1983boo,Pyboo} for particle $i$,
\begin{align}
	q_2(i) =& \sqrt{\frac{4\pi}{5} \sum_{m=-2}^{2} |q_{2,m}(i)|^2 }, \label{eq:ql}\\
	q_{2,m}(i) =& \frac{1}{N_i}\sum_{j=1}^{N_i} Y_{2,m}(\theta(\mathbf{r}_{ij}),\phi(\mathbf{r}_{ij})),
	\label{eq:qlm}
\end{align}
where the $Y_{\ell, m}$ are spherical harmonics and $\mathbf{r}_{ij}$ is one of the $N_i$ bonds involving particle $i$.

\subsubsection{Fourier space analysis}
Our experimental data do not have periodic boundary conditions, so we must use a window function to ensure the correct correlation, especially at small $q$. Here we use the Hanning window, that significantly affects only the values of $S(q)$ at the first lowest five $q$ that we discard in the rest of the analysis. We checked that our results are not affected by other reasonable choices of the window function.

The main wavevector is defined as
\begin{equation}
<q> = \frac{\int_{0}^{q_\mathrm{min}}dq\, q\, S(q)}{\int_{0}^{q_\mathrm{min}}dq\, S(q)},
\end{equation}
where $q_\mathrm{min}$ is fixed at all times at a value that corresponds to the minimum between the low-q peak and the hard sphere peak.

\begin{acknowledgments}
This study was partly supported by Grants-in-Aid for Scientific Research (S) (21224011) and Specially Promoted Research (25000002) from the Japan Society of the Promotion of Science (JSPS).
Collaboration between M.L. and H.Tanaka has been funded by CNRS through Projet international de coopération scientifique No~7464. 
M.L. acknowledges support from ANR grant GelBreak ANR-17-CE08-0026. 
J.R. acknowledges support from the European Research Council Grant DLV-759187 and the Royal Society University Research Fellowship.
\end{acknowledgments}

\bibliographystyle{apsrev4-1}
\bibliography{Gelation_Mechanics}



\section*{Supplementary material}

# Supplementary information for
# Gelation as condensation frustrated by hydrodynamics and mechanical isostaticity

Hideyo Tsurusawa,[1, *] Mathieu Leocmach,[2, *] John Russo,[3] and Hajime Tanaka[4, †]

[1] Department of Fundamental Engineering, Institute of Industrial Science,
University of Tokyo, 4-6-1 Komaba, Meguro-ku, Tokyo 153-8505, Japan
[2] Univ Lyon, Université Claude Bernard Lyon 1, CNRS,
Institut Lumière Matière, F-69622, VILLEURBANNE, France
[3] School of Mathematics, University of Bristol, Bristol BS8 1TW, United Kingdom
[4] Institute of Industrial Science, University of Tokyo,
4-6-1 Komaba, Meguro-ku, Tokyo 153-8505, Japan

**Supplementary Analysis of isostaticity percolation**

We compute the cluster size distribution of all particles at percolation time, and also the isostatic cluster size distribution at the percolation time of isostaticity. We take into account the finite size of the particles by adding one particle radius to the radius of gyration $R_g$ of each cluster. The normalized cluster size is thus

$$\frac{R_g}{\sigma} + \frac{1}{2} = \frac{1}{\sigma}\sqrt{\frac{1}{s}\sum_{i=1}^{s}\left|\vec{X}_i - \frac{1}{s}\sum_{i=1}^{s}\vec{X}_i\right|^2} + \frac{1}{2} \qquad (1)$$

where $s$ is the cluster size. In that way a one particle cluster has a size of $0.5\sigma$. In that way the fractal range is extended to small clusters. Figure S6(a) show the resulting cluster size distributions for a dilute gel. At usual percolation time, the central range of cluster sizes exhibit a fractal dimension compatible with diffusion-limited cluster aggregation ($D = 1.85$). Large clusters that have an extent comparable to the experimental window display a compact ($D = 3$) fractal dimension. At small cluster sizes we also observe compactness, consistent with the scenario where compaction at small scales proceeds before diffusion-limited percolation. This signature of the original compaction remains when isostaticity percolates. However larger isostatic clusters exhibit a fractal dimension more compatible with directed percolation ($D = 2.27$). This is consistent with the simultaneity of the isotropic percolation of isostaticity and the directed percolation that is observed in dilute samples, see Fig. 5(a) of the main text. At higher volume fraction, Figure S6(b) shows that usual percolation is just random ($D = 2.53$) at all cluster sizes. However when isostaticity percolates the cluster size distribution of isostatic clusters displays a fractal dimension compatible with directed percolation ($D = 2.27$). Since we know that directed percolation of the whole network took place before isostaticity percolation, see Fig. 5(b), this observation is consistent with a simple invasion of the network by isostaticity.

To get more insight on the way directed percolation and isostaticity percolation are related in the dilute case, we take a look at what changes as isostaticity invades the network. We take as reference configuration the percolation time $t_{\text{perco}}$. For low volume fraction gels, local compaction has already occurred at that time, and we detect hundreds of small isostatic clusters that are embedded in a percolated but non isostatic network. We define $\mathcal{X}_i(t)$ the position of the center of mass at time $t$ of the set particles that formed isostatic cluster $i$ at $t = t_{\text{perco}}$. We note $\mathcal{X}_{ij}(t) = |\mathcal{X}_j(t) - \mathcal{X}_i(t)|$ the Euclidian distance between the centers of mass of $i$ and $j$, and $\Delta\mathcal{X}_{ij}(t) = \mathcal{X}_{ij}(t) - \mathcal{X}_{ij}(t_{\text{perco}})$ its increment. When taking an ensemble average over all pairs of clusters that are not directly connected or far apart in the network $\langle\Delta\mathcal{X}_{ij}(t)\rangle$ is null. However when considering only pairs of clusters that are less than 10 bonds away on the network, we can observe a shortening of the distance between them, see Fig. 6(e). It means that, as loose strands are converted to isostaticity, see Fig. 6(b)-(d), these strands shorten and straighten. This change in morphology allows longer directed path and thus promotes directed percolation.

---

[*]These authors contributed equally to this work
[†]Electronic address: tanaka@iis.u-tokyo.ac.jp

## Supplementary Figures

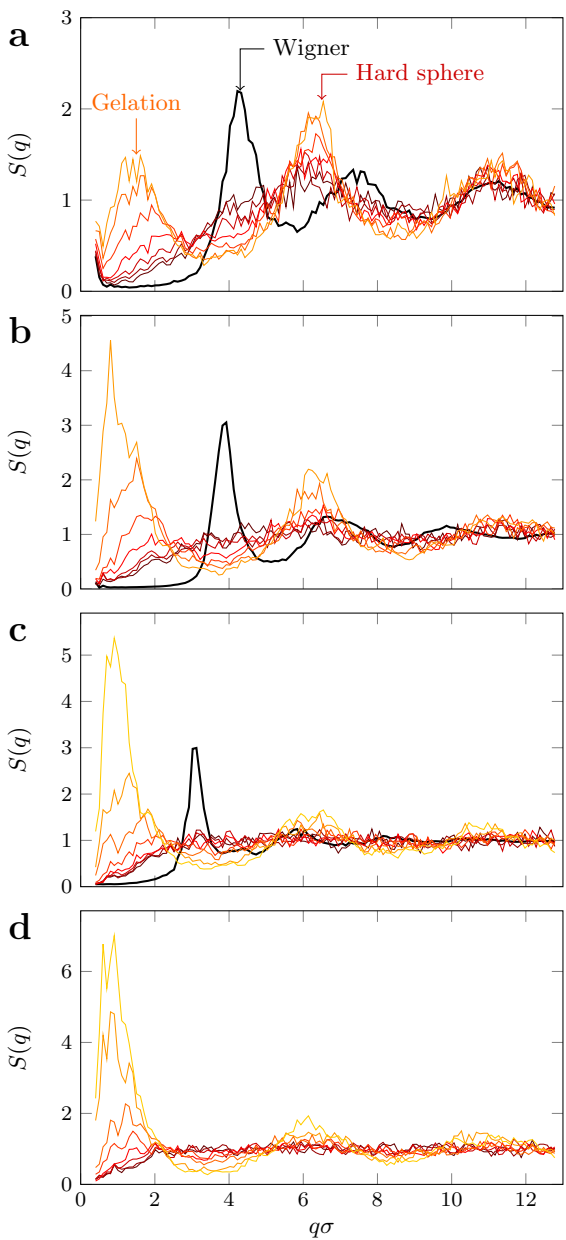

Fig. S1: Temporal change of the structure factor. Panels (a)-(d) are for the four samples shown on Fig. 3 of the main text by decreasing volume fraction. The thick black curve corresponds to the initial Wigner crystal before salt introduction (ill defined thus not shown in (d)). Thin curves from dark red to yellow are spaced by 150 s and display a peak corresponding to the hard sphere diameter as well as a growing peak at low $q$ indicating gelation.

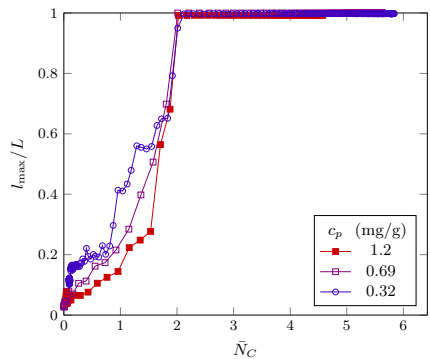

Fig. S2: Gelation path dependence on polymer concentration. Comparison of system evolution in terms of largest cluster extent and of mean coordination number. By decreasing polymer concentration: $\phi = 16, 15, 14$ % and $c_p = 1.2, 0.69, 0.32$ mg/g for ■, □ and ○ respectively.

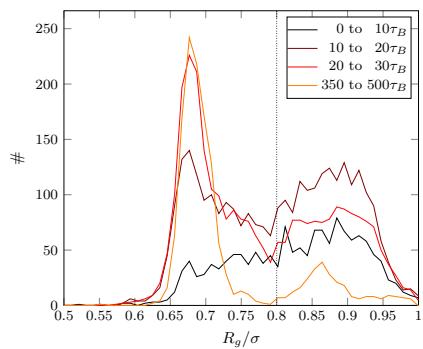

Fig. S3: Evolution of the population of triplets as a function of their radius of gyration. The result is for a non percolating sample ($\phi = 4$ %, $c_p = 1$ mg/g).

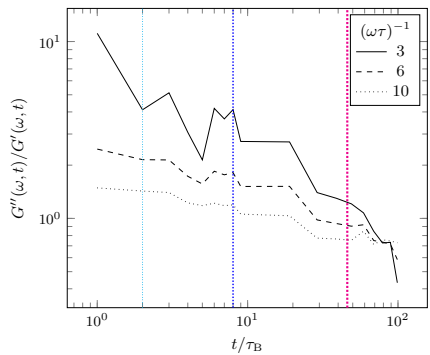

Fig. S4: Loss tangent. Evolution of the ratio $G''(\omega,t)/G'(\omega,t)$ for three frequencies for the sample at $\phi = 27$ %, $c_p = 1$ mg/g. The dotted vertical lines show the percolation time of particles having at least 2, 4 and 6 neighbors respectively of particles having at least 2, 4 and 6 neighbors respectively, from left to right.

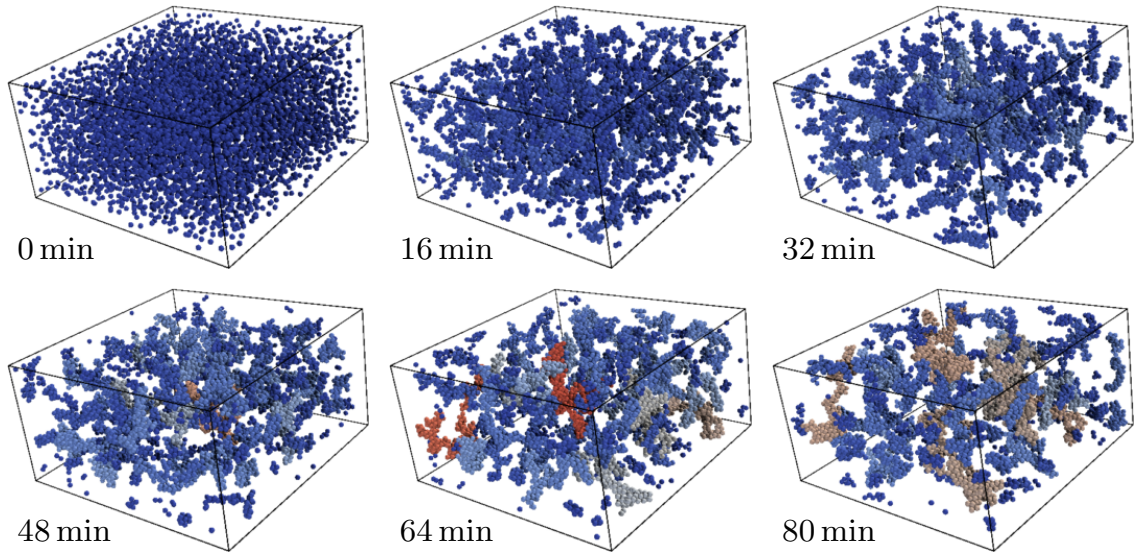

Fig. S5: Cluster phase formation observed by our method. Experimental coordinates are reconstructed and colored by the number of particles in the cluster ($\phi = 4.7$ %, $c_p = 1$ mg/g).

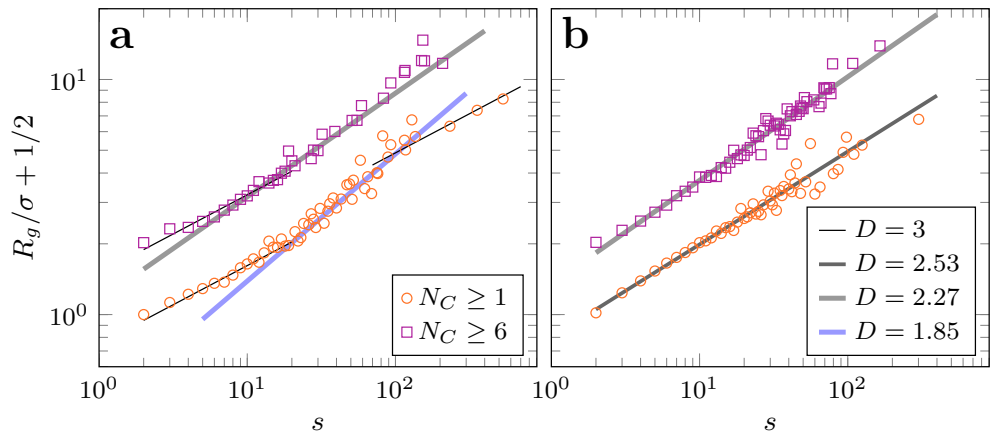

Fig. S6: Cluster size distributions considering either all particles (orange) or isostatic particles (purple), at their respective isotropic percolation time. The later is shifted by a factor 2 for clarity. The thin black lines correspond to a fractal dimension of 3 (compact), dark grey lines to a fractal dimension of 2.53 (random percolation), thick light grey lines to a fractal dimension of 2.27 (directed percolation), and thick light blue lines correspond to a fractal dimension of 1.85 (DLCA). (a) $\phi = 8$ %, $c_p = 1.5$ mg/g. (b) $\phi = 27$ %, $c_p = 1$ mg/g.

**Supplementary Movies**

**Supplementary Movie 1.** Reconstructions in a thin slice from confocal coordinates of the whole process of gelation at $\phi = 7.7$ %, $c_p = 1$ mg/g. The phase separation is induced by salt injection. Particles are drawn to scale, and colored according to the radius of gyration of the cluster they belong to. Time stamp is $(t - t_0)/\tau_B$. We can see that there is little macroscopic flow in this protocol.

**Supplementary Movie 2.** Reconstructions in a thin slice from confocal coordinates of the whole process of gelation at $\phi = 27$ %, $c_p = 1$ mg/g. The depth of view is 7.5 $\mu$m, while the lateral dimension is 140 $\mu$m. Isostatic particles are drawn to scale, non isostatic ones are drawn smaller for clarity. The bond network is displayed in orange. Time stamp is $(t - t_0)/\tau_B$. In this sample, isotropic percolation occurs at $t = 2\tau_B$, directed percolation at $4\tau_B$ and isostaticity percolation at $46\tau_B$.


\end{document}